# THE RADIATIVE $^{10}$Be(n,γ)$^{11}$Be CAPTURE AT THERMAL AND ASTROPHYSICAL ENERGIES

## Dubovichenko S.B.,[1] Dzhazairov-Kakhramanov A.V.[2]

*V.G. Fessenkov Astrophysical Institute, National Center of Space Research and Technologies, ASC MID RK, Almaty, Kazakhstan*
[1]*dubovichenko@mail.ru*, [2]*albert-j@yandex.ru*

**Abstract**
Within the framework of the modified potential cluster model with the forbidden states and the classification of states according to Young tableaux the possibility of describing the available experimental data for the total reaction cross sections of the neutron radiative capture on $^{10}$Be at thermal and astrophysical energies was shown.

## 1. Introduction

In the middle of nineties in papers [1,2] the possibility of describing the Coulomb form factors of $^{6,7}$Li, having a high degree of clusterization in the two-body channels, based on the single-channel two-body potential cluster model (PCM) was demonstrated [3-5]. This variant of the model is based on the concept of forbidden states (FSs) [4-9] in the intercluster potentials. The presence of the FSs leads to the appearance of "extra" nodes in the wave functions (WFs) of the relative cluster motion. The forbidden states were used by us more in papers [10,11], the consideration of which in a simple PCM allowed one to get an acceptable description of the main characteristics of some light nuclei. Furthermore, in the PCM with the consideration of tensor forces [12,13], the possibility of a correct reproduction of almost all main characteristics of $^6$Li was demonstrated, including the well-known, at that time, the value of its quadrupole moment.

Furthermore, in papers [3-5,14-33] the possibility of an acceptable description of the known experimental data on the astrophysical *S*-factors or the total cross sections for radiative capture on some light nuclei was demonstrated. Namely, the capture in systems p$^2$H, n$^2$H, p$^3$H, p$^6$Li, n$^6$Li, p$^7$Li, n$^7$Li, n$^8$Li, p$^9$Be, n$^9$Be, p$^{10}$B, n$^{10}$B, p$^{11}$B, n$^{11}$B, p$^{12}$C, n$^{12}$C, p$^{13}$C, n$^{13}$C, p$^{14}$C, n$^{14}$C, n$^{14}$N, p$^{15}$N, n$^{15}$N, n$^{16}$O and $^2$H$^4$He, $^3$He$^4$He, $^3$H$^4$He, $^4$He$^{12}$C at thermal and astrophysical energies was considered. The calculations of these capture processes are based on a modified version of PCM with the FSs and the classification of cluster states according to Young tableaux (MPCM), the methods of which were described in details in papers [26-33] and [34].

This quite remarkable success of the MPCM can be explained by the fact that the potentials of intercluster interaction in the continuous spectrum are constructed not only on the basis of the known elastic scattering phase shifts, but also taking into account the classification of cluster states according to Young tableaux [35]. The potential parameters of the bound states (BSs) at a given number of the allowed states (ASs) and forbidden states in the given partial wave are also fixed quite clearly. For this purpose, for instance, the requirement of description of the nucleus binding energy in the cluster channel and its charge radius and two-body asymptotic constant (AC) is used [34]. Furthermore, such potentials allow one to perform the calculations, for example, of the astrophysical *S*-factors of the radiative capture reactions [36] or the total cross sections of these reactions [37] at low and ultralow energies. These calculations include, in some cases the thermal energy range and as a whole allow to reproduce the available experimental data for the total cross sections of the capture reactions and some characteristics of the BSs for the most considered light nuclei [3,34].

Continuing the study of radiative capture processes [3,34], let's consider the reaction n+$^{10}$Be → $^{11}$Be+γ in the frame of the MPCM at thermal and astrophysical energies. This reaction is a part of one of the variant of the chain of primordial nucleosynthesis of the Universe [38]

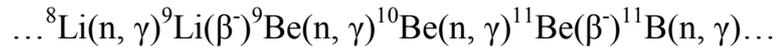

…$^{8}$Li(n, γ)$^{9}$Li(β$^{-}$)$^{9}$Be(n, γ)$^{10}$Be(n, γ)$^{11}$Be(β$^{-}$)$^{11}$B(n, γ)…

by which the elements with a mass of *A* > 11-12 can be formed (see., e.g., [39]).

## 2. The structure of the n$^{10}$Be states

For $^{10}$Be, as well as for $^{10}$B [34], we accept the Young tableau {442}, therefore for the n$^{10}$Be system we have {1} × {442} → {542} + {443} + {4421} [35,40]. The first from the obtained tableaux compatible with the orbital angular moments *L* = 0, 2, 3, 4, and is forbidden because it is forbidden to have more than four nucleons in the *s*-shell, the second tableau is allowed and compatible with the orbital angular moments *L* = 1, 2, 3, 4, and the third is also allowed, compatible with *L* = 1, 2, 3 [35].

Generally speaking, the lack of the tables of products of the Young tableaux for the number of particles 10 and 11 makes it impossible to make an accurate classification of the cluster states in the given system of particles. However, even such a qualitative estimation of orbital symmetries allows one to determine the presence of the FSs in the *S* wave and lack of the FSs for the *P* states. Exactly such structure of the FSs and ASs in the different partial waves allows to build further the potentials of intercluster interaction, which are necessary for the calculation of the total cross sections for the considering radiative capture reaction [3,34].

Thus, taking into account only the lower partial waves with orbital angular moments *L* = 0, 1, 2, it can be said that for the n$^{10}$Be system (for $^{10}$Be $J^{\pi}$,*T* = 0$^{+}$,1 [41]) in the potentials of the $^{2}P$ waves only the allowed state presents, and $^{2}S$ and $^{2}D$ waves have the forbidden states. The state in the $^{2}S_{1/2}$ wave (in the notation of $^{(2S+1)}L_J$), corresponds to the GS of $^{11}$Be with $J^{\pi}$,*T* = 1/2$^{+}$,3/2 and is located at the binding energy of the n$^{10}$B system of -0.5016 MeV [42].

Note that the $^{2}P$ waves correspond to the two allowed Young tableaux {443} and {4421}. This situation seems to be similar to the systems N$^{2}$H or N$^{10}$B, described in [3,14,28-31,34], where the potentials for the scattering processes depend on two Young tableaux, and for the BS only on one [43,44]. Therefore, here we assume that the potential $^{2}P_{1/2}$ of the BS (first excited state – FES) corresponds to one tableau {443}. Consequently, the potentials of the $^{2}P_{1/2}$ BS and of the $^{2}P_{1/2}$ scattering processes are different, because they depend on a different set of Young tableaux. To fix the idea we will assume that for a discrete spectrum the allowed state in the $^{2}P_{1/2}$ wave is bound, while for the scattering processes it is not bound. Therefore, the depth of such potential can be simply set equal to zero. The FS occurs to be the bound state for the $^2S_{1/2}$ scattering wave or for the discrete spectrum in the n$^{10}$B system.

Now let us consider the FES, bound in the n$^{10}$Be channel, and the first resonance state (FRS) of $^{11}$Be [42], which is not bound in the n$^{10}$Be channel and corresponds to the resonance in the n$^{10}$Be scattering. The FES of $^{11}$Be is located at energy of 0.32004 MeV comparatively to the GS with the $J^\pi = 1/2$ moment or -0.18156 MeV comparatively to the n$^{10}$Be channel threshold. This state can be related to the doublet $^2P_{1/2}$ level without FS. The first resonance state is located at 1.783 MeV comparatively to the GS or at 1.2814 MeV comparatively to the n$^{10}$Be channel threshold. For this level the moment $J^\pi = 5/2^+$ [42] is provided, which allows to take $L = 2$ for it, i.e., to consider it as the $^2D_{5/2}$ resonance in the n$^{10}$Be system at 1.41 MeV in the laboratory system (l.s.), and its potential has the FS. The width of such resonance is equal to 100(10) keV in the center of mass (c.m.) [42].

On the basis of these data, it is reputed that the $E1$ capture from the $^2P$ scattering waves with the potential of zero depth without the FSs to the $^2S_{1/2}$ GS of $^{11}$Be with the bound FS is possible.

No. 1. $\begin{array}{c} ^2P_{1/2} \to ^2S_{1/2} \\ ^2P_{3/2} \to ^2S_{1/2} \end{array}$.

For the radiative capture to the FES the similar $E1$ transition from the $^2S_{1/2}$ and $^2D_{3/2}$ scattering waves with the bound FSs to the $^2P_{1/2}$ BS without FS is possible.

No. 2. $\begin{array}{c} ^2S_{1/2} \to ^2P_{1/2} \\ ^2D_{3/2} \to ^2P_{1/2} \end{array}$.

The GS potentials and the FES will be constructed further in a way for correct description of the channel binding energy, the charge radius of $^{11}$Be and its asymptotic constant in the n$^{10}$Be channel. Therefore, the known values of the asymptotic normalization coefficient (ANC) and the spectroscopic factor $S$, with help of which the AC is found, have a quite big error, the GS potentials will also have several variants with different parameters of width, which strongly affect the value of the AC.

The data on the asymptotic normalization coefficients $A_{NC}$, for example, are given in paper [48]. Here we will also use the well-known relation

$$A_{NC}^2 = S \times C^2, \qquad (1)$$

where $S$ – the spectroscopic factor, $C$ – asymptotic constant in fm$^{-1/2}$, which is connected with the dimensionless AC [45] $C_W$, used by us as follows: $C = \sqrt{2k_0} C_W$, and the dimensionless constant $C_W$ is given by the expression [45]

$$\chi_L(r) = \sqrt{2k_0} C_W W_{-\eta L+1/2}(2k_0 r), \qquad (2)$$

where $\chi_L(r)$ – numerical wave function of the bound state, obtained from the solution of the radial Schrödinger equation and normalized to unity, $W_{-\eta L+1/2}$ – the Whittaker function of the bound state, which determines the asymptotic behavior of the wave function and is a solution of the same equation without nuclear potential, $k_0$ – the wave number caused by the $E$ channel energy $k_0 = \sqrt{2\mu \frac{m_0}{\hbar^2} E}$, $\eta$ – the Coulomb parameter $\eta = \frac{\mu Z_1 Z_2 e^2}{\hbar^2 k}$, determined numerically $\eta = 3.44476 \cdot 10^{-2} \frac{\mu Z_1 Z_2}{k}$ and $L$ – the orbital angular moment of the bound state. Here $\mu$ – the reduced mass of particles of the input channel, and the constant $\hbar^2/m_0$ was assumed equal to 41.4686 fm$^2$, where $m_0$ is the atomic mass unit (amu).

In further calculations we used the radius of $^{10}$Be in the GS equals 2.357(18) fm from paper [46], and for the GS of $^{11}$Be the known radius value of 2.463(15) fm is given in [42]. The charge radius of the neutron assumed to be zero, and its massive range 0.8775(51) fm coincides with the known radius of the proton [50]. In addition, for the charge radius of the FES of $^{11}$Be the calculated value of 2.43(10) fm [47] is known, and for the GS the value 2.42(10) fm is obtained in the same paper. For the radius of the neutron in $^{11}$Be estimation equals 5.6(6) fm is given [47]. At the same time in paper [48] for the neutron radius in the GS the value of 7.60(25) fm is given, while for the FES of -4.58(25) fm. In the all calculations for the masses of nucleus and neutron the exact values are used: $m(^{10}\text{Be}) = 10.013533$ amu [49] and $m(\text{n}) = 1.00866491597$ amu [50].

## 3. Methods for calculating of the total cross sections

The total cross sections of the radiative capture $\sigma(NJ,J_f)$ for the $EJ$ transitions in the potential cluster model are given, for example, in [3,28-31,34] or [36] and have the form

$$\sigma_c(NJ, J_f) = \frac{8\pi K e^2}{\hbar^2 q^3} \frac{\mu}{(2S_1+1)(2S_2+1)} \frac{J+1}{J[(2J+1)!!]^2} A_J^2(NJ, K) \cdot$$

$$\cdot \sum_{L_i, J_i} P_J^2(NJ, J_f, J_i) I_J^2(J_f, J_i),$$

where $\sigma$ – the total cross section of the process of the radiative capture, $\mu$ – the reduced mass of particles of the input channel, $q$ – the wave number of particles of the input channels, $S_1$, $S_2$ – spins of the particles in the input channel, $K$, $J$ – the wave number and the moment of $\gamma$ quantum in the output channel, $N$ – it is the $E$ or $M$ transitions of $J$-th multipolarity from the initial state $J_i$ to the final state $J_f$ of the nucleus.

For the electric orbital $EJ(L)$ transitions ($S_i = S_f = S$) the value $P_J$ has the form [3,34]

$$P_J^2(EJ, J_f, J_i) = \delta_{S_i S_f}[(2J+1)(2L_i+1)(2J_i+1)(2J_f+1)](L_i 0 J 0 | L_f 0)^2 \begin{Bmatrix} L_i & S & J_i \\ J_f & J & L_f \end{Bmatrix}^2$$

$$A_J(EJ, K) = K^J \mu^J \left( \frac{Z_1}{m_1^J} + (-1)^J \frac{Z_2}{m_2^J} \right), \qquad I_J(J_f, J_i) = \langle \chi_f | R^J | \chi_i \rangle \qquad (3)$$

Here $S_i$, $S_f$, $L_f$, $L_i$, $J_f$, $J_i$ are the total spins and moments of particles in the input (*i*) and the output (*f*) channel; μ, $m_1$, $m_2$, $Z_1$, $Z_2$ are the reduced mass, the masses and the charges of particles in the input channel; $I_J$ is the integral over the wave functions of the initial $\chi_i$ and final $\chi_f$ state as the functions of the relative motion of clusters, in this case, the neutron and $^8$Li, with the interparticle distance $R$.

For the spin part of the magnetic process $M1(S)$ in the MPCM the following expression is used ($S_i = S_f = S$, $L_i = L_f = L$) [34]:

$$P_1^2(M1, J_f, J_i) = \delta_{S_i S_f} \delta_{L_i L_f} \left[ S(S+1)(2S+1)(2J_i+1)(2J_f+1) \right] \begin{Bmatrix} S & L & J_i \\ J_f & 1 & S \end{Bmatrix}^2,$$

$$A_1(M1, K) = i \frac{\hbar K}{m_0 c} \sqrt{3} \left[ \mu_1 \frac{m_2}{m} - \mu_2 \frac{m_1}{m} \right], \quad I_J(J_f, J_i) = \langle \chi_f | R^{J-1} | \chi_i \rangle, \quad J=1. \quad (4)$$

Here $m$ – mass of the nucleus, $\mu_1$, $\mu_2$ – the magnetic moments of the clusters, the other symbols are the same as in the previous terms. For the neutron magnetic moment and $^{10}$Be the following values were used: $\mu_n = -1.91304272\mu_0$ and $\mu(^{10}\text{Be}) = 0$ [41], where $\mu_0$ – nuclear magneton. The correctness of the above expression for the $M1$ transition pre-tested in our previous papers [3,14,17,34] on the basis of neutron radiative capture on $^7$Li and proton radiative capture on $^2$H reactions at low and astrophysical energies.

## 4. The n$^{10}$Be interaction potentials

As usual [3,14-17,26-31,34], in the capacity of the n$^{10}$Be interaction in each partial wave with a given orbital angular moment $L$ we use the potential of the Gaussian form with the point-like Coulomb term

$$V(r, L) = -V_L \exp(-\gamma_L r^2). \qquad (5)$$

The ground state of $^{11}$Be in the n$^{10}$Be channel is the $^2S_{1/2}$ level and this potential should describe the AC of this channel correctly. In order to extract this constant from the available experimental data, let us consider information about the spectroscopic factor $S$ and the asymptotic normalization coefficients $A_{NC}$. The results for $A_{NC}$ are given in paper [48] that are presented in Table 1, here some results from paper [38] are added. In addition to this, a relatively large amount of data for the spectroscopic factors of the n$^{10}$Be channel of $^{11}$Be is managed to find [42], so we give their values in the separate Table 2.

Table 1. The $A_{NC}$ data of $^{11}$Be in the n$^{10}$Be channel

| Reaction from which the $A_{NC}$ is determined | The value of the $A_{NC}$ in fm$^{-1/2}$ for the GS | The value of the $A_{NC}$ in fm$^{-1/2}$ for the FES | Literature |
|---|---|---|---|
| (d,p$_0$) at 12 MeV | 0.723(16) | 0.133(4) | [48] |
| (d,p$_0$) at 25 MeV | 0.715(35) | 0.128(6) | [48] |
| (d,p$_0$) at 25 MeV | 0.81(5) | 0.18(1) | [38] |
|  | 0.68–0.86 | 0.122–0.19 | *Interval* |
|  | *0.749* | *0.147* | *Average $\overline{A}_{NC}$* |

Furthermore, based on the expression (1) for the GS we find $\overline{A}_{NC}/\sqrt{\overline{S}} = \overline{C} = 0.94$ fm$^{-1/2}$, and since $\sqrt{2k_0} = 0.546$, then the dimensionless AC defined as $\overline{C}_W = \overline{C}/\sqrt{2k_0}$, is equal to $\overline{C}_W = 1.72$. However, the range of values of the spectroscopic factor is so high that the value $C_W$ may be in the range of 1.54-2.29, and if we consider the errors of $A_{NC}$, then this interval can be extended to 1.40-2.63. For the FES at $\sqrt{2k_0} = 0.423$ we find $\overline{C}_W = 0.45$ similarly, and the range of values $\overline{C}_W$ for the average ANC is equal to 0.35-0.62. If we consider the $A_{NC}$ errors, then this interval is expanded to 0.29-0.81.

Table 2. Data for the spectroscopic factors $S$ of $^{11}$Be in the n$^{10}$Be channel

| The $S$ value for the GS | The $S$ value for the FES | Literature |
|---|---|---|
| 0.42(6) | 0.37(6) | [51] |
| 0.72(4) | --- | [52] |
| 0.61(5) | --- | [53] |
| 0.56(18) | 0.44(8) | [54,55] |
| 0.73(6) | 0.63(15) | [56,57] |
| 0.77 | 0.96 | [58] |
| *0.36–0.79* | *0.31–0.96* | *Interval* |
| *0.64* | *0.6* | *Average $\overline{S}$* |

The potential $^2S_{1/2}$ of the GS with the FS, which allows one to get the dimensionless constant $C_W$, close to the average value of 1.72, has the parameters

$$V_{1/2} = 47.153189 \text{ MeV and } \gamma_{1/2} = 0.1 \text{ fm}^{-2}. \tag{6}$$

It leads to the binding energy of -0.501600 MeV with an accuracy of the used herein finite-difference method (FDM) for the calculating of the binding energy of 10$^{-6}$ MeV [7], the AC $C_W = 1.73(1)$ on the interval of 7-30 fm, the mass radius of 3.16 fm, the charge radius of 2.46 fm. The determination of the estimated expressions of these radiuses is given, for example, in papers [1-3,34]. The AC errors are determined by its averaging over the specified range of distances.

Such potential of the GS with the FS is in a full accordance with the classification of states according to Young tableaux given above, and gives the charge radius of $^{11}$Be which is in a good agreement with data [42]. The parameters of the GS potential were constructed on the basis of the approximate description of the average value of the AC equals 1.72 received above, and its phase shift that shown in figure 1 by the solid line. This potential at the orbital angular moment $L = 2$ leads to the non-resonant $^2D$ scattering phase shift without spin-orbital splitting shown in figure 1 by the dotted line. On the same figure the $^2S_{1/2}$ phase shifts of the n$^{10}$Be scattering, obtained in the calculations in work [59], are shown by points.

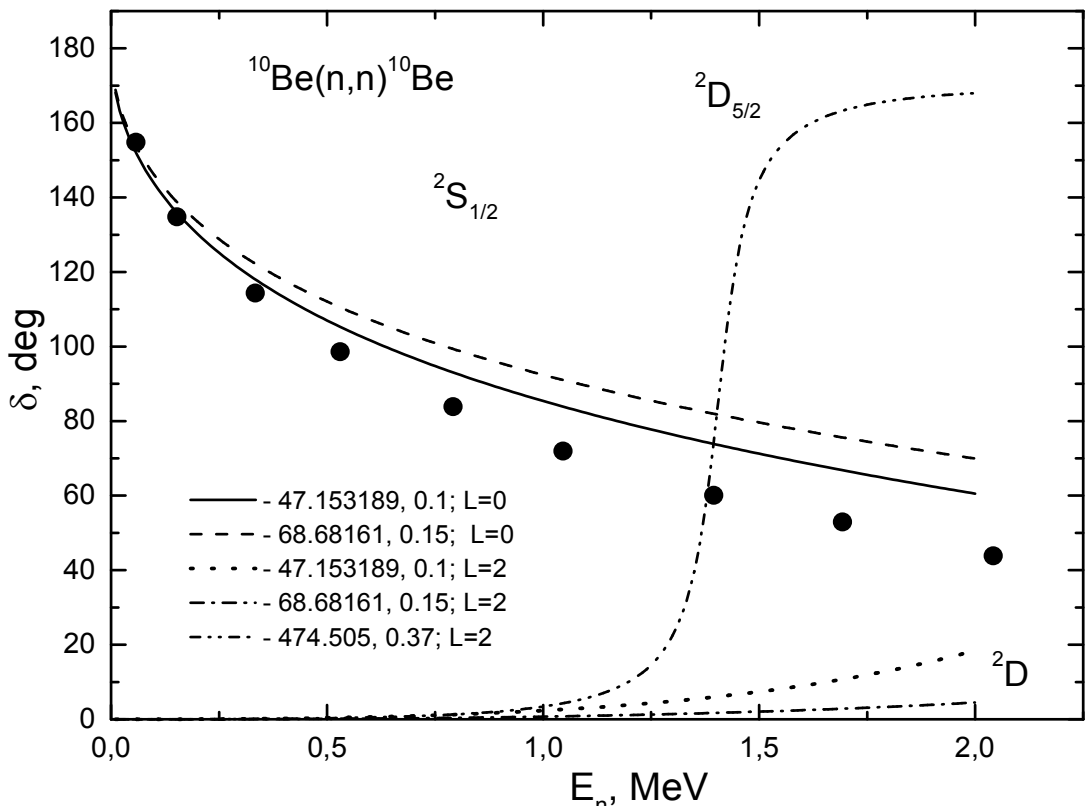

Figure 1. The n$^{10}$Be elastic scattering phase shifts of the $^2S_{1/2}$ wave. The meaning of the lines and points is explained in the text.

To compare the results, let us consider another variant of the potential of the GS with the FS and the parameters

$$V_{1/2} = 68.68161 \text{ MeV and } \gamma_{1/2} = 0.15 \text{ fm}^{-2}. \qquad (7)$$

The potential leads to the binding energy of -0.501600 MeV at the same accuracy of the FDM, $C_W = 1.56$ (1) on the interval of 5–30 fm, the mass radius of 3.05 fm, the charge radius of 2.45 fm and its phase shift is shown in figure 1 by the dashed line. The dimensionless AC is close to the lower limit of the range of values of this magnitude 1.54–2.29. Such potential with the orbital angular moment $L = 2$ leads to the $^2D$ scattering phase shift without spin-orbital splitting, shown in figure 1 by the dash-dotted line.

Let's note that for the potential of the resonance $^2D_{5/2}$ wave with the FS, which will be required further for the consideration of the $E2$ transitions the following parameters were obtained

$$V_{5/2} = 474.505 \text{ MeV and } \gamma_{5/2} = 0.37 \text{ fm}^{-2}. \qquad (8)$$

The potential leads to the resonance at 1.41 MeV (l.s.) at width of $\Gamma_{c.m.} = 100$ keV, that is fully consistent with the data [42], and its phase shift is shown in figure 1 by the dot-dot-dashed line.

The potential $^2P_{1/2}$ of the FES without the FS can have the parameters

$$V_{1/2} = 9.077594 \text{ MeV and } \gamma_{1/2} = 0.03 \text{ fm}^{-2}. \qquad (9)$$

The potential leads to the binding energy of -0.181560 MeV with an accuracy of the FDM of $10^{-6}$ MeV [7], the AC equals 0.73(1) on the interval of 11–30 fm, the mass radius of 3.58 fm and the charge radius of 2.52 fm. The phase shift of such potential decreases gradually and at 2.0 MeV has the value, approximately, up to 115°. The potential parameters of the FES (9) were chosen for the correct description of the total cross sections of the neutron capture on $^{10}$Be at thermal energy of 25.3 MeV, obtained in paper [38], and the value of its dimensionless AC is in the above allowable range of 0.29–0.81 values.

Now we return to the criteria of constructing potentials for the $^2P$ scattering wave, which may differ from the potential of the FES because of the different Young tableaux of these states [43,44]. First of all, as was indicated above, such potential should not have the forbidden state. Therefore, we do not have the results of the phase shift analysis of the n$^{10}$Be elastic scattering, and in the spectra of $^{11}$Be at energies below 2.0 MeV there are no resonances of the negative parity, we will consider that the $^2P$ potentials should lead in this energy range to almost zero scattering phase shifts – so they simply may have the null depth. For the potential of the $^2S_{1/2}$ scattering the interaction $^2S_{1/2}$ of the GS with the FS will be used, for instance, the variant of the potential (6), because it leads to the relatively good agreement with the scattering phase shifts from paper [59], shown by the solid line and the points in figure 1.

## 5. The total cross sections for the neutron radiative capture on $^{10}$Be

As already mentioned, we assume that the $E1$ radiation capture No. 1 comes from the $^2P$ scattering wave to the $^2S_{1/2}$ GS of $^{11}$Be in the n$^{10}$Be channel. Our calculations of such capture cross sections for the potential of the GS (6) lead to the results, shown in figure 2 by the dashed line, and the results for the potential of the GS (7) are presented by the solid line. In all of these calculations for the $^2P$ elastic scattering potentials the potential of zero depth was used. The experimental data of the neutron radiative capture on $^{10}$Be are shown in figure 2 by points and are given in paper [60] with reference to the paper [61]. As seen from these results, the calculated cross sections describe the available experimental data in a relatively narrow energy range, approximately from 0.3–0.4 MeV to 2.0 MeV. The calculated line decreases rapidly at lower energies and doesn't describe the data for thermal energy, as shown in figure 3.

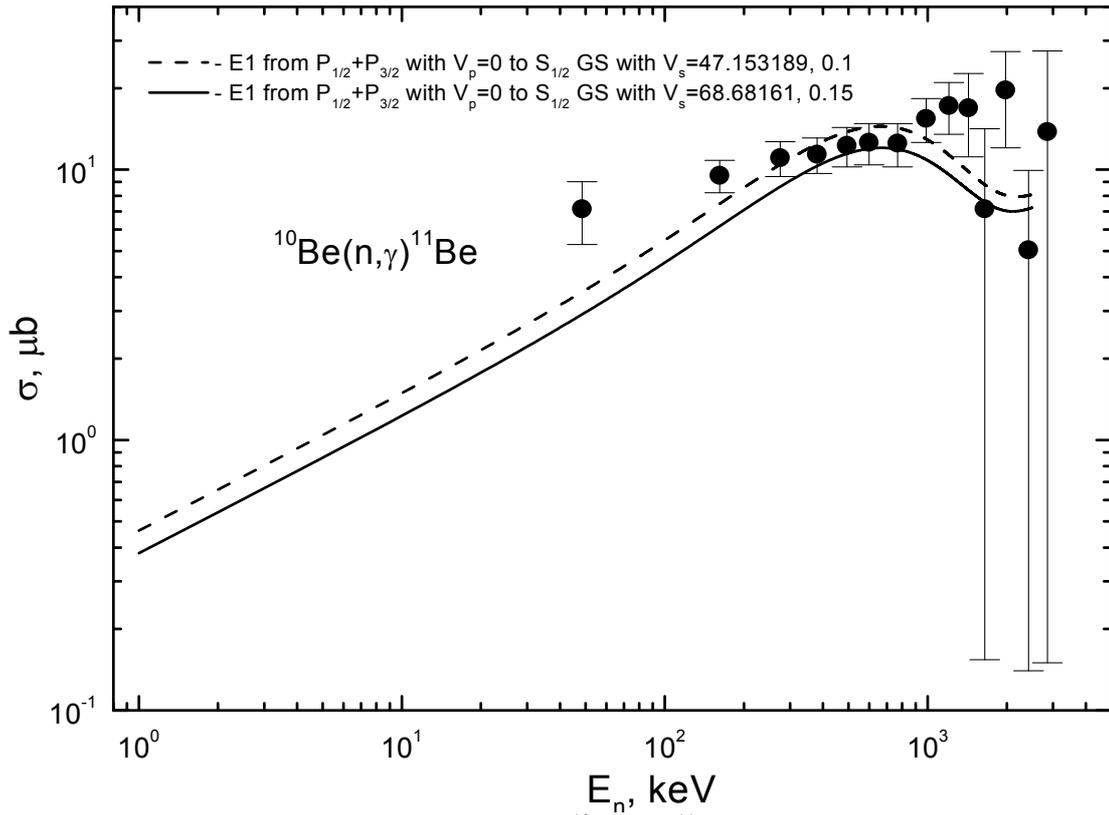

Figure 2. The total cross sections of the radiative $^{10}\text{Be}(n,\gamma)^{11}\text{Be}$ $E1$ capture to the GS. Points are the experimental data from paper [60]. The meaning of lines is explained in the text.

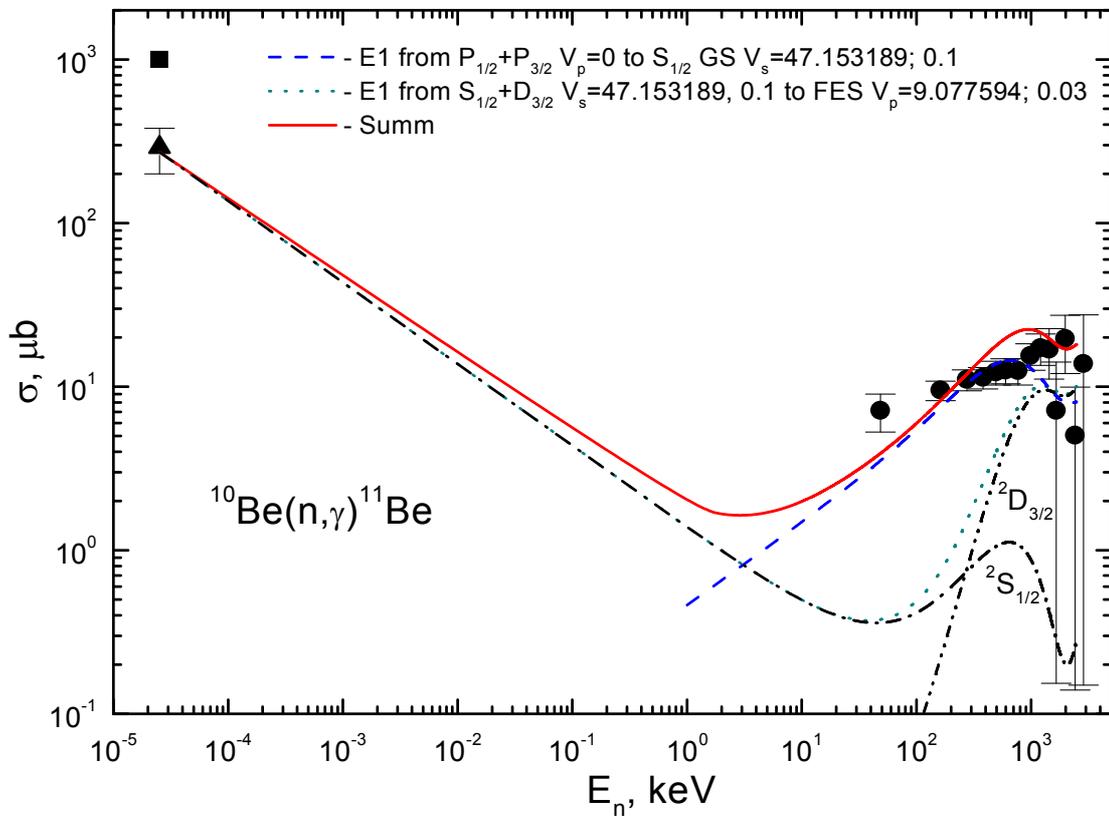

Figure 3. The total cross sections of the radiative $^{10}\text{Be}(n,\gamma)^{11}\text{Be}$ capture. The experimental data of papers: [61] – points, [38] – triangle, [62] – square. The meaning of lines is explained in the text.

Therefore, further we will consider the $E1$ transitions to the $^2P_{1/2}$ FES from the $^2S_{1/2}$ and $^2D_{3/2}$ scattering wave. The results for the $E1$ transition to the GS with the potential (6) and the $^2P$ scattering potential are still shown in figure 3 by the dashed line. The cross sections for the $E1$ transition No. 2 from the $^2S_{1/2}+^2D_{3/2}$ scattering wave with the potential (6) for $L = 0$ and 2 to the $^2P_{1/2}$ FES with the potential (9) are shown by the dotted line. The solid line shows the summarize cross section of these two $E1$ processes, which, in the large, reproduces the general course of the available experimental data correctly in the given energy region – from the thermal 25.3 meV to 2.0 MeV.

The contribution of the transition from the $^2S_{1/2}$ wave is shown by the dot-dashed line and from the $^2D_{3/2}$ scattering wave by the dot-dot-dashed line. All coefficients in the expression (3) were calculated for the $^2D_{3/2}$ wave, and the $^2S_{1/2}$ wave potential (6) used in these calculations has the orbital angular moment $L = 2$, i.e., corresponds to the $^2D$ state without the spin-orbital. The calculated cross section at thermal energy was found to be 272 μb. The experimental data at thermal energy were taken from paper [38] – triangle with the value of 290(90) μb and [62] – the square, which indicates the upper limit of the thermal cross section equals of 1 mb.

For the comparison, now we use the potential of the GS with parameters (7). These results with the same potential of the FES (9) are shown in figure 4 (notation as in figure 3), with a cross section at thermal energy equals of 343 μb. It is evident that such potential of the GS is slightly better describes the cross section at energies from 0.1-0.2 to 2.0 MeV. Thus, the variants of the calculations with the potentials of the GS (6) and (7) and the potentials of the FES (9) lead to the general description of the available data at the energy of 25.3 meV and in the range of about 0.1–2.0 MeV. The 50–100 keV energy range is described comparatively bad, so let us consider the other variants of the transitions at the neutron capture on $^{10}$Be.

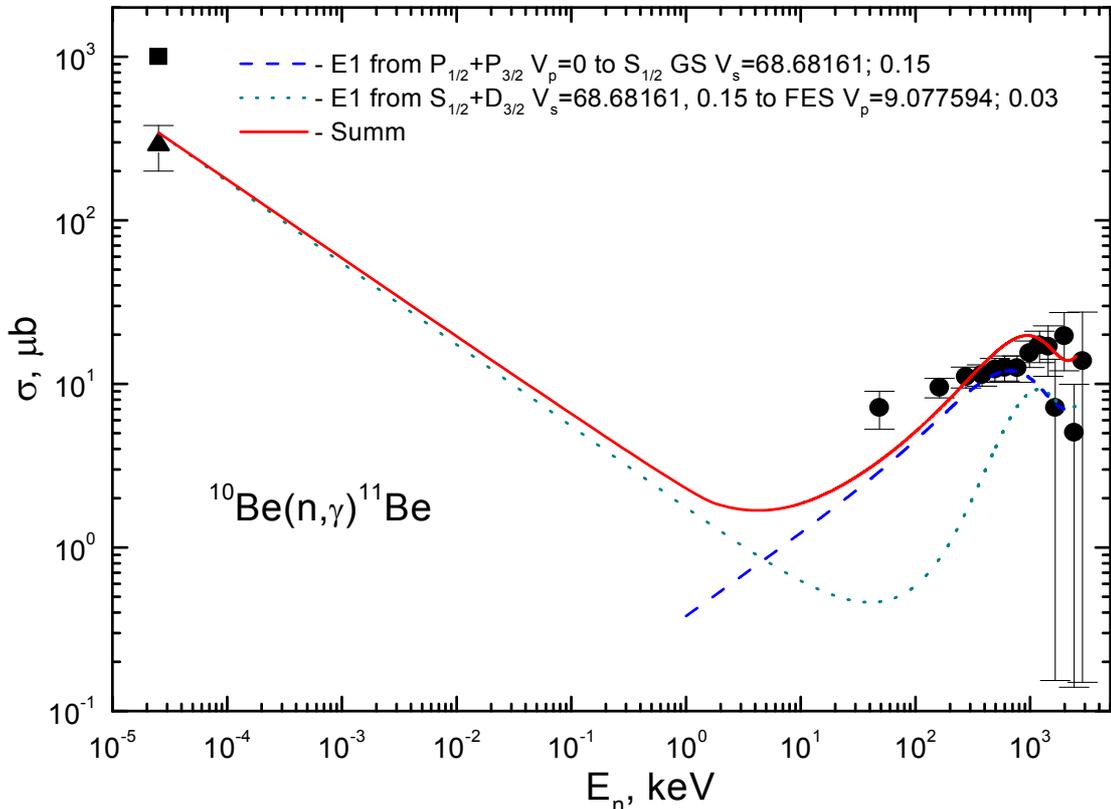

Figure 4. The total cross sections of the radiative $^{10}$Be(n,γ)$^{11}$Be capture. The experimental data of papers: [61] - points, [38] - triangle, [62] - square. The meaning of lines is explained in the text.

The cross section of the possible $M1$ transition from the $^2S_{1/2}$ scattering wave to the $^2S_{1/2}$ GS of $^{11}$Be in the n$^{10}$Be channel with the same potential (6) or (7) in the both states will tend to zero because of the orthogonality of the wave functions of discrete and continuous spectra in the same potential. The actual numerical calculation of these cross sections leads to a value less than $10^{-2}$ μb in the energy range from 1 keV to 2.0 MeV, and at the energy of 25.3 meV the cross section occurs to be slightly less than 1% from the cross section of the transition to the FES, shown in figure 4 by the dotted line.

If we consider the $M1$ transitions from the $^2P$ scattering waves with the zero potential to the $^2P_{1/2}$ FES with the potential (9), then the sections don't exceed 0.15 μb in the entire energy region. For the $E2$ transitions from the $^2D_{3/2}$ wave with the potential (6) or (7) at $L = 2$ and the $^2D_{5/2}$ wave with the potential (8) to the GS with the $^2S_{1/2}$ even at the resonance energies the value of these cross sections do not exceed $10^{-3}$ μb. Hence, it is clear that such transitions are not contribute significantly to the total cross section of the considered process and the problem of description of the cross sections in the range from 50 keV to 100–200 keV remains open.

Since at energies of 25.3 meV and up to about 10 eV, the calculated cross section is a straight line (solid line in figure 4), it can be approximated by a simple function of energy of the form [34]

$$\sigma_{ap}(\mu b) = \frac{A}{\sqrt{E_n(\text{keV})}}. \quad (10)$$

The value of the constant $A = 1.7265$ μb·keV$^{1/2}$ was determined by a single point in the calculated cross section at minimum energy equals of 25.3 meV. The module of the relative deviation

$$M(E) = \left|[\sigma_{ap}(E) - \sigma_{theor}(E)]/\sigma_{theor}(E)\right| \quad (11)$$

of the calculated theoretical cross section ($\sigma_{theor}$) and approximation ($\sigma_{ap}$) of this cross section of the given above function (10) in the energy range up to 10 eV is located at the level of 0.1%.

It is realistic to assume that this form of dependence of the total cross section from energy will also be saved at lower energies. Therefore, based on the given expression (10) for approximation of the cross section one can perform the estimation of the cross section, for instance, at energy of 1 μeV (1 μeV = $10^{-9}$ keV), which gives a value of about 54.6 mb.

## 6. Conclusion

Thus, in the frame of the MPCM with the classification of states according to Young tableaux is quite possible to construct the potentials of the n$^{10}$Be interaction, which allow one to reproduce the general course of the experimental data for the total cross sections of the radiative neutron capture on $^{10}$Be at low and ultralow energies generally correct. The theoretical cross sections are calculated from thermal energy 25.3 meV to 2.0 MeV and approximated by a simple function of energy, which can be used to calculate the cross sections at energies below 10–50 eV.

The proposed variants of the potentials of the ground and the first excited states of $^{11}$Be in the n$^{10}$Be channel allow to obtain the AC within available for it errors, and lead to a reasonable description of the $^{11}$Be radius. The result shows that in the 29th cluster system of light nuclei in the framework of the single-channel MPCM it is possible to describe some of the main characteristics of the nucleus and, in general, the total cross section of the neutron radiative capture on $^{10}$Be correctly.

## Acknowledgments

This work was supported in the framework of the grant program "Studying of the thermonuclear processes in the Universe" of the Ministry of Education and Science of the Republic of Kazakhstan through the V.G. Fessenkov Astrophysical Institute, "NCSRT" ASC MID RK.

In conclusion, the authors express their deep gratitude to Strakovsky I.I. (GWU, Washington, USA) and Uzikov Y.N. (JINR, Dubna, Russia) for a discussion of certain issues touched in the paper.